# An Interactive 3D Visualization Tool for Large Scale Data Sets for Quantitative Atom Probe Tomography

Hari Dahal,[1] Michael Stukowski,[2] Matthias J. Graf,[3] Alexander V. Balatsky[1] and Krishna Rajan[2]

[1]Theoretical Division and Center for Integrated Nanotechnology, Los Alamos National Laboratory, Los Alamos, NM 87544
[2]Materials Science and Engineering & Institute for Combinatorial Discovery, Iowa State University, Ames, IA 50011
[3]Theoretical Division, Los Alamos National Laboratory, Los Alamos, NM 87544

## ABSTRACT

Several visualization schemes have been developed for imaging materials at the atomic level through atom probe tomography. The main shortcoming of these tools is their inability to parallel process data using multi-core computing units to tackle the problem of larger data sets. This critically handicaps the ability to make a quantitative interpretation of spatial correlations in chemical composition, since a significant amount of the data is missed during subsequent analysis. In addition, since these visualization tools are not open-source software there is always a problem with developing a common language for the interpretation of data. In this contribution we present results of our work on using an open-source advanced interactive visualization software tool, which overcomes the difficulty of visualizing larger data sets by supporting parallel rendering on a graphical user interface or script user interface and permits quantitative analysis of atom probe tomography data in real time. This advancement allows materials scientists a codesign approach to making, measuring and modeling new and nanostructured materials by providing a direct feedback to the fabrication and designing of samples in real time.

## INTRODUCTION

The current state-of-the art Atom Probe Tomography (APT) hardware comes with its own visualization tool, e.g., Imago visualization and analysis system (IVAS) [1]. There exist several other visualization tools that have been developed by different atom probe experimental groups, e.g. ADAM [2]. The main shortcoming of these tools is their inability to parallel process data using multi-core computing units to tackle the problem of large data sets. This severely limits the data analysis and leads to analyzing only parts of the data at a time. Analyzing only parts of the data normally is fraught with missing important information. This dilemma can be overcome using a codesign approach to materials design, fabrication, characterization, and analysis by developing state-of-the-art visualization and analysis capabilities, which offer a streamlined feedback to the making, measuring and modeling of materials in real time.

In addition, since these visualization tools are not open-source software there is always a problem with developing a common language for the interpretation of data. In our analysis we use an open-source software package 'ParaView', which overcomes the difficulty of visualizing larger data sets by supporting parallel rendering in GUI (Graphical User Interface) and SUI (Script User Interface) for batch jobs. Even though this tool was not developed specifically to analyze atom probe data, one can show that it can do most of the qualitative and quantitative analysis performed using advanced analysis methods. Additionally it provides an efficient platform for analyzing larger data sets and the development of custom modules.

The information on ParaView software can be obtained through the website: http://paraview.org/. ParaView is a general-purpose visualization and advanced analysis tool for a wide array of data. It can provide an alternative platform for the analysis of scientific data irrespective of their origin [3-7]. It runs on distributed and single node architectures, in parallel and serial configurations, as well as with distributed or shared memory.

Some of the advanced visualization capabilities integrated in this tool are: interactivity and scalability, support for multiple platforms and data formats, availability of more than fifty filters to apply on the data, availability of both GUI and SUI environments, ability to extract quantitative information, support for active stereo visualization etc. Interactivity provides an opportunity to interact with the data in real time. Scalability eliminates the limit of the data size one can work with. The support for multiple data formats makes this tool useful for data coming from different experiments. The availability of many filters makes it useful for cross-analysis of the data. Some of the most common and useful filters are, just to name a few: glyphs, calculator, plot variable along a line, plot dynamic variable in three dimensional space, 3D Delaunay, Fourier transformation, statistical analysis, extraction of subset of data, analysis of a part of the data without actually restructuring the data, histogram, contours and isosurfaces, analysis of vector fields, masking, transparency and many more [8]. Several of these filters help to improve the qualitative analysis of the data, in addition, many of them are quite useful in extracting the quantitative information from the data. The script user interface is particularly useful in analyzing large data sets through batch jobs. Scripting is also quite useful to extend the standard capabilities of the visualization tool to meet the user's specific needs, because one can integrate suitable algorithms into the main tool in a modular fashion. We would like to point out that ParaView is not the only open-source software package that has these capabilities and versatility; for example, VisIt (https://wci.llnl.gov/codes/visit/) is equally capable. Moreover, we do not envision ParaView as a substitute for the existing commercial and/or open-source software, but a complement to the exiting capabilities.

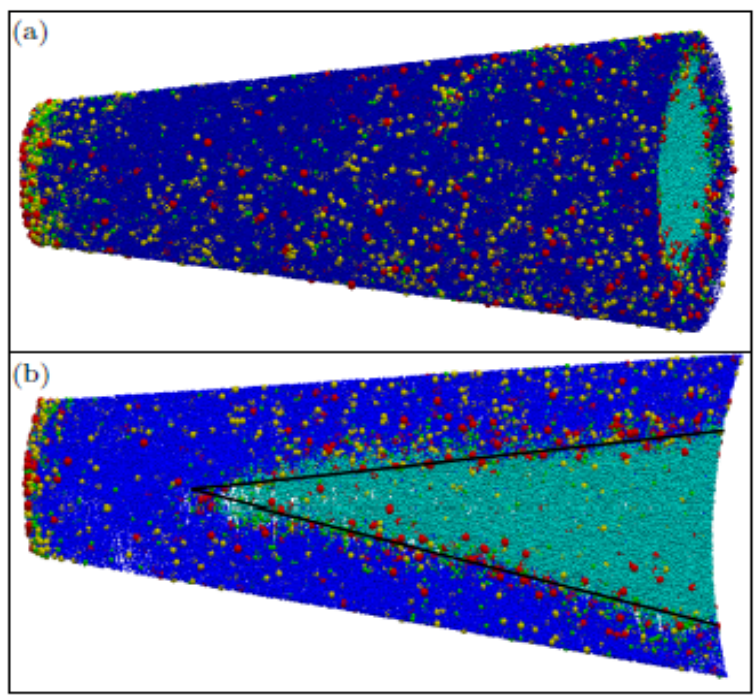

FIG. 1. The color of the glyphs represents the constituent elements: blue (Cu), turquoise (Si), green (O), yellow (CuO), and red (SiO). The figure is constructed using the 'Sphere' glyph with a linear scaling of the radius of the sphere for different element type. The APT tip has a thin layer of oxide (left side of the tip). The structure has a distinct boundary between the Cu and Si region.

**RESULTS AND DISCUSSION**

For the purpose of the presented visualization and advanced data analysis study, we used a simple example for proof-of-principle of the codesign approach in materials design. We chose a sample with test geometry of an APT silicon tip coated with a Cu film to help easily delineate a clear compositional discontinuity. This test geometry was useful to experiment with. It provided a diversity of imaging and analysis issues: identification and delineation of a chemical composition interface, detection of minority elements as well as oxide dimers. All this information is crucial for the design and fabrication of novel and nanostructured materials.

We would like to demonstrate here that the need for using parallel rendering can originate due to a very simple problem. Using a two dimensional (2D) glyph is always computationally less expensive, but in our study we find that it is very difficult to distinguish the constituent elements of a three dimensional (3D) object due to the ultra-dense nature of the data set. Using the 3D sphere glyph to visualize the distribution of elements provided the solution to this problem, since it allows an emphasis of minority elements, however, it poses other challenges. When we use the 2D vertex glyph the total memory required is 56MB; a single-processor computer can be used to render the entire data. However when we use the sphere glyph (with 8 points of polar angle $\theta$ resolution and 8 points of azimuthal angle $\phi$ resolution) a standard single-processor unit approaches its memory and CPU (Central Processing Unit) limitations. In this case, processing times of graphical operations became the bottleneck and we could visualize only a small subset of randomly selected data points using the 'Masking' filter provided in ParaView. Alternatively, we could visualize only a small portion of the sample. None of these cases was convenient for interactive or real time data analysis. This situation demanded a need to use an advanced visualization tool that can support parallel rendering. Paraview is capable of parallel rendering, as is VisIt. Hence we rendered the object in a small computing cluster with distributed memory. We used ParaView in Client-Server mode in which the parallel rendering is performed on the compute cluster (source) and the data is analyzed on the desktop with a high-end graphics card (client). We would like to note here that we checked the scalability of ParaView for our data set. We used 500,000 data points to be represented by sphere glyphs having 64-polygons per sphere. The time to render a single image varied linearly with the number of processors used (2, 4, 8, 16, 32 processors take 138, 70, 35, 18, 9 seconds respectively). From these tests it is clear that if further interactive manipulations of the 3D object are required it would be impractical to achieve it with a single-processor computer.

We have developed a module for ParaView that can be used for complex data manipulation in interactive or background mode to facilitate the materials design, fabrication and characterization aspect. Our developed modular script permits the dynamic interrogation of chemical composition along the Z-axis of the specimen tip. As mentioned above, the visualization tool we are using allows just that using both the GUI and SUI environment. We can define a rectangular clip box with a fixed volume and change its position by specifying a variable 'Position' along the Z-axis in the GUI of ParaView. It allows us to analyze the distribution of the elements in the clipped region anywhere on the Z-axis. In fact the change in the position of the box-cut can be automated using the 'Animation view' filter provided in the GUI and in a script using the SUI. We start out with a clipped box perpendicular to the Z- axis at z=0 and change its position in steps of 1 nm along the Z-axis. We can automate the change in the position of the clipped region and record the distribution of the elements in each clipped region as shown in Fig. 2b. It is also possible to save the created animation both as a movie and image sequence. By visually analyzing the image sequence or the animated movie we see that the boundary between the Cu and Si rich regions is well separated throughout the entire atom probe tip.

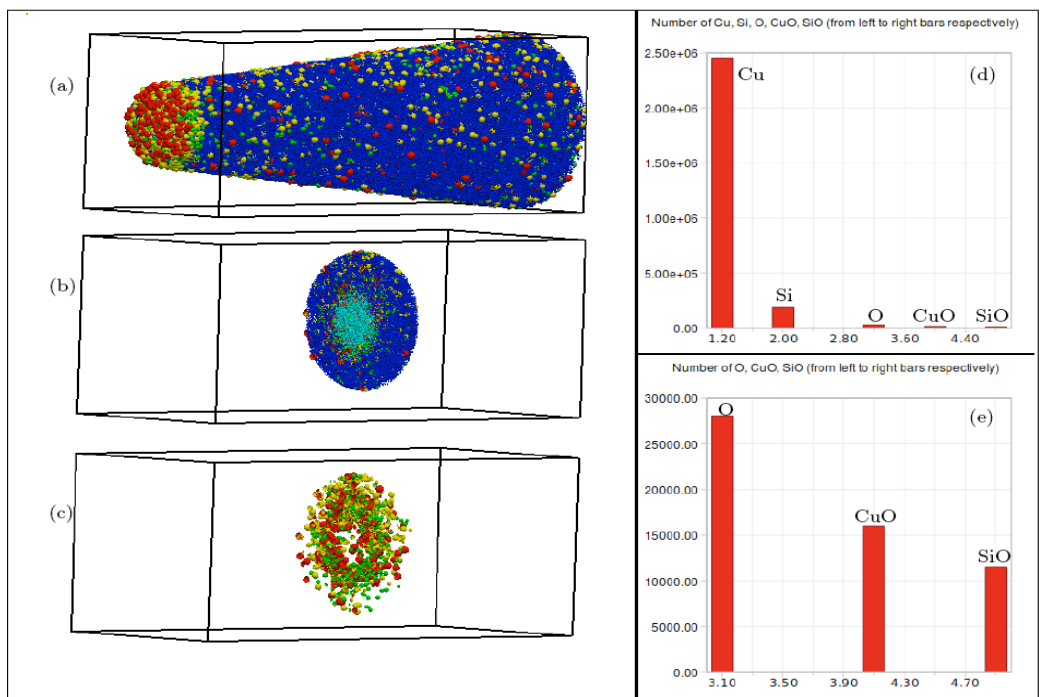

FIG. 2: Spatial distribution and histogram of all elements and only the oxidized elements in a small cross-sectional clipped region at z=50nm are shown. a) A transversal box cut of the width 1 nm along the Z-axis at z=50nm b) All the constituent elements in this clipped region are shown. The clipped region is separated from the mother figure by translating it down using the 'Translation' filter. c) Only oxygen and the oxide dimers are filtered out using the 'Threshold' filter. d) The histogram of all the constituent elements obtained by using 'Histogram' filter on the clipped region is shown. The number on the Y-axis is scaled by the number of polygons used to create a sphere to represent one element (particle). The number gives the relative abundance of the elements in the region. e) The histogram of only the oxygen and oxide dimers are shown. The whole image is a single screen shot of the GUI.

**CONCLUSIONS**

We have developed new visualization and advanced materials analysis capabilities based on an open-source software package (ParaView), which can be used in a modular approach. These new capabilities augment existing commerical packages, thus aiding the visualization and data analysis of large data sets of atom probe images. We developed a script-based algorithm for parallel use in the client-server model as a modular implementation in ParaView. In this paper, we focused on the demonstration of a highly interactive visualization platform for quantitative analysis of chemical composition within the specimen. It provided new insights into materials properites at the nano scale and the compositional interfaces important for materials design. It shows the value of parallel rendering of 3D objects to efficiently analyze large data sets in real time, and suggests a solution to this problem in terms of already existing visualization software, where customized analysis modules can be added by the user. The use of such open-source software can in general be applied to the visualization and analysis challenges of large data sets from atom probe tomography.  We envisoned the application of advanced visualization tool as an integral part of materials codesign since it provides a real time feedback to the materials fabrication and characterization.

**ACKNOWLEDGMENTS**

We are indebted to James Ahrens and John Patchett for their help in using the ParaView visualization software. This work was carried out under the auspices of the National Nuclear Security Administration of the U.S. Department of Energy at Los Alamos National Laboratory under Contract No. DE-AC52-06NA2539 and Office of Science (BES). KR and MS acknowledge support from the National Science Foundation: NSF-CDI Type II program: grant no. PHY 09-41576 and NSF-AF grant no. CCF09-17202, NSF-ARI Program: CMMI 09-389018. This work was also supported in part by the Defense Advanced Research Projects Agency (DARPA) under Grant N66001-10-1-4004, N/MEMS Fundamental Science and Technology Research Centers Program, Dr. T. Akinwande, Program Manager and AFOSR (Grant:FA0550-10-1-0256). KR would also like to acknowledge support from Iowa State University through the Wilkinson Professorship in Interdisciplinary Engineering.

**REFERENCES:**

1. http://www.imago.com/imago
2. O. Hellman, J. Vandenbroucke, J. Blatz du Rivage, D. N. Seidman, Materials Science and Engineering A 327, 29 (2002)
3. M. J. Graf, J. Ahrens, J. Patchett, H. Dahal, A. V. Balatsk, D. Modl, L. Monroe, N. Brown, E. Akhadov, SciDAC Review, Issue 10, Winter 2008, 32 (2008)
4. J. Ahrens, K. Heitmann, S. Habib, L. Ankeny, P. McCormick, J. Inman, R. Armstrong and K.L. Ma, Journal of Physics: Conference Series - Scientific Discovery Through Advanced Computing, Volume 46, 526 (2006)
5. http://www.paraview.org/paraview/resources/applications.html
6. http://supercomputing.iu.edu/sc2005/applications/physical/paraview_visl.php
7. http://www-vis.lbl.gov/NERSC/Software/paraview/
8. http://www.paraview.org/paraview/help/download/ParaView%20User's%20Guide%20v3.10.pdf